\documentstyle[aps,psfig,twocolumn]{revtex}
\begin{document}
\flushbottom
\twocolumn[
\hsize\textwidth\columnwidth\hsize\csname @twocolumnfalse\endcsname

\title{Gapped tunneling spectra in
the normal state of Pr$_{2-x}$Ce$_x$CuO$_4$}
\author{Amlan Biswas$^1$, P. Fournier$^2$, V. N. Smolyaninova$^1$, 
R. C. Budhani$^1$,
J. S. Higgins$^1$ and R.~L.~Greene$^1$}
\address{1. Center for Superconductivity Research, 
Department of Physics, University of Maryland,
College~Park, MD-20742}
\address{2. Centre de Recherche sur les Propri\'et\'es \'electroniques
de mat\'eriaux avanc\'es, D\'epartement de Physique, Universit\'e
de Sherbrooke, Qu\'ebec, CANADA J1K 2R1}
\date{\today}
\maketitle
\tightenlines
\widetext
\advance\leftskip by 57pt
\advance\rightskip by 57pt

\begin{abstract}
We present tunneling data in the normal state of the electron doped
cuprate superconductor Pr$_{2-x}$Ce$_x$CuO$_4$ for three different
values of the doping $x$. The normal state is obtained by applying
a magnetic field greater than the upper critical field, $H_{c2}$ for
$T < T_c$.
We observe an anomalous normal state gap near the Fermi level. From our
analysis of the tunneling data we conclude that this is a feature
of the normal state density of states. We discuss possible reasons for
the formation of this gap and its implications for the nature of the
charge carriers in the normal and the superconducting states of cuprate
superconductors.
\end{abstract}
\pacs{}
]
\narrowtext
\tightenlines

\section{Introduction}

The nature of the normal state of high-$T_c$ superconductors is a
matter of great interest because it gives
information about the nature of the charge carriers
and the pairing mechanism leading to superconductivity. 
It is now widely accepted that, in cuprate superconductors
there is a depletion of the
density of states (DOS) at the Fermi level ($E_F$) at a 
temperature $T^*$, which is much higher than the superconducting
transition temperature ($T_c$). This gap is called the pseudogap (PG).
The evidence for this PG has come from various experiments which 
include angle resolved photoemission spectroscopy (ARPES), 
optical conductivity and tunneling spectroscopy ~\cite{suzuki,timusk}. 
For the hole-doped high-$T_c$ superconductors (e.g. BSCCO) these
measurements show that the superconducting gap ($\Delta_{SC}$)
has a similar width as the PG. This leads many to believe that the
PG and the $\Delta_{SC}$ have a similar origin. Theories predicting
pair correlations above $T_c$ have been suggested, which are
supported by these experiments. The other important property of the
PG is its doping dependence. The PG appears at higher temperatures
for the underdoped compounds and is not observed in overdoped
compounds ~\cite{timusk}.
The situation is less clear for other cuprates like the
hole-doped system La$_{2-x}$Sr$_x$CuO$_4$ (LSCO).
The width of the PG in LSCO from
ARPES ~\cite{sato} and optical conductivity ~\cite{startseva}
measurements comes out to be much larger than the width of 
$\Delta_{SC}$ ~\cite{alff2}. 
These results
raise questions about the origin of the PG and more experiments on the
normal state of the cuprates are necessary. 

A lot of interest has been generated recently by the electron doped
counterparts of the high $T_c$ 
cuprates. These materials have the general 
formula $RE_{2-x}$Ce$_x$CuO$_4$ ($RE$=Nd, Pr, Sm).
These materials were believed to have an $s$-wave 
superconducting gap symmetry
but recent results
have shown that the gap has a $d$-wave symmetry like the hole-doped
cuprates ~\cite{patgap,prozorov,kirtley}.
A comparison between the properties of the hole-doped and
electron doped systems is needed as input for theories of high-$T_c$
superconductivity.
In this context it is important to investigate the normal state of the
electron-doped
materials and check if results similar to the hole-doped cuprates are
obtained, e.g. the formation of a PG in the normal
state.
Transport measurements have shown that there is a metal-insulator 
crossover as a function of Ce doping
in the normal state of the electron-doped cuprate Pr$_{2-x}$Ce$_x$CuO$_4$
(PCCO) ~\cite{pat1}.
Recently tunneling spectroscopy
measurements have been reported in the normal state of PCCO and NCCO
~\cite{ourpap,alff1}
These reports give evidence for a normal state gap
for fields above the bulk $H_{c2}$ and $T < T_c$. 
In ref. ~\cite{alff1} the authors suggest
that this normal state gap could be due to the presence of a PG in the normal
state of the electron doped superconductors. However the width of this
gap is not consistent with other recent measurements using ARPES and optical
conductivity which show normal state gaps of 60 meV and wider 
~\cite{armitage,basov2}. 
Further work is therefore
necessary to explain the origin of this normal state gap and
the differences between the 
properties of the hole and electron-doped systems. 

In this paper we report our tunneling studies of the normal state of
the electron doped cuprate superconductor
Pr$_{2-x}$Ce$_x$CuO$_4$ for different values of the doping $x$.
Preliminary work was reported in ~\cite{ourpap}.
For PCCO the $H_{c2}$ is of the order of 10 
tesla for $x=0.15$ at 1.5 K. This makes
it possible to obtain the normal state by applying a field greater
than $H_{c2}$ and to then
perform the tunneling studies. In hole-doped cuprates the $H_{c2}$
is much higher ($\ge$ 50 T for optimally doped samples)
and the tunneling studies which have been performed in the normal
state are for $T > T_c$. We find that in the 
normal state, there is a depletion of the density of states (DOS) 
near the Fermi level ($E_F$). Further investigation reveals that
this normal state gap (NSG) closely 
resembles the correlation gap formed in disordered metals due to 
enhanced electron-electron interactions. We present data which show
that this NSG is not due to residual superconductivity at the surface
for fields above the bulk $H_{c2}$ and we argue that this is
probably not a pseudogap feature. 
The formation of this NSG gives us
new information about the nature of the charge carriers in the normal
state and hence about the pairing mechanism.

There have been several tunneling studies reported on the  electron-doped
cuprates but the results have been inconsistent {\em vis a vis}
the values of the SC gap, the observation of the zero bias 
conductance peak formed
due to Andreev bound states, etc ~\cite{huang,yamamoto,ekino}. 
The standard techniques used for
other cuprates like YBCO, e.g. multilayer junctions, are difficult to
use for the electron doped materials because, 1) a-axis oriented 
superconducting films of PCCO or NCCO have not been grown yet, and 2)
the extreme sensitivity of these compounds to the oxygen stoichiometry
which makes it difficult to fabricate tunnel junctions without changing
the superconductor in some way. In this paper we report tunneling data
on three different types of junctions and show that they all give
consistent results.

\section{Experimental Details}

The tunneling studies were performed mainly on $c$-axis oriented
thin films of PCCO
grown using pulsed laser deposition (PLD) on LaAlO$_3$ (LAO)
and yttrium stabilized zirconia (YSZ) substrates. Details of the film growth
are given in ref. ~\cite{patfilm1,patfilm2}. 
The films have been optimized for oxygen content by maximizing $T_c$
for each cerium concentration. The films were characterized by
X-ray, ac susceptibility and resistivity measurements.

We have used three techniques to tunnel into the $a-b$ plane
of $c$-axis oriented thin films.
In the first technique, a thin film of c-axis oriented
PCCO, covered with SiO$_2$, 
is broken in vacuum to expose the a-b plane while Ag is being
evaporated in the vacuum chamber. A junction is thus formed between
the freshly exposed PCCO surface and silver. The size of the junction
is reduced to about 200 $\mu$m. This is a new method of forming junctions
to tunnel into the a-b plane of PCCO details of which will be discussed
in a future publication.
We will call these ``break junctions'' in 
the subsequent sections although they are not break junctions in the
usual sense. 
The second technique is very similar to the break junction method. Here
the c-axis oriented film is covered with SiO$_2$ and broken in air.
Indium was pressed on the freshly
exposed side of the film. We used indium to check the quality of the
junction by recording the tunneling features of indium.
These will be called ``pressed junctions''.
The third method was the usual point contact junction formed by pressing
a gold electrode on the side of a single crystal or a freshly broken thin
film such that the direction of current flow is perpendicular to the 
c-axis.

\section{Results and Discussions}

We first show the tunneling data on the optimally doped PCCO sample
($x=0.15$). Fig. 1a shows 
\begin{figure}
\centerline{
\psfig{figure=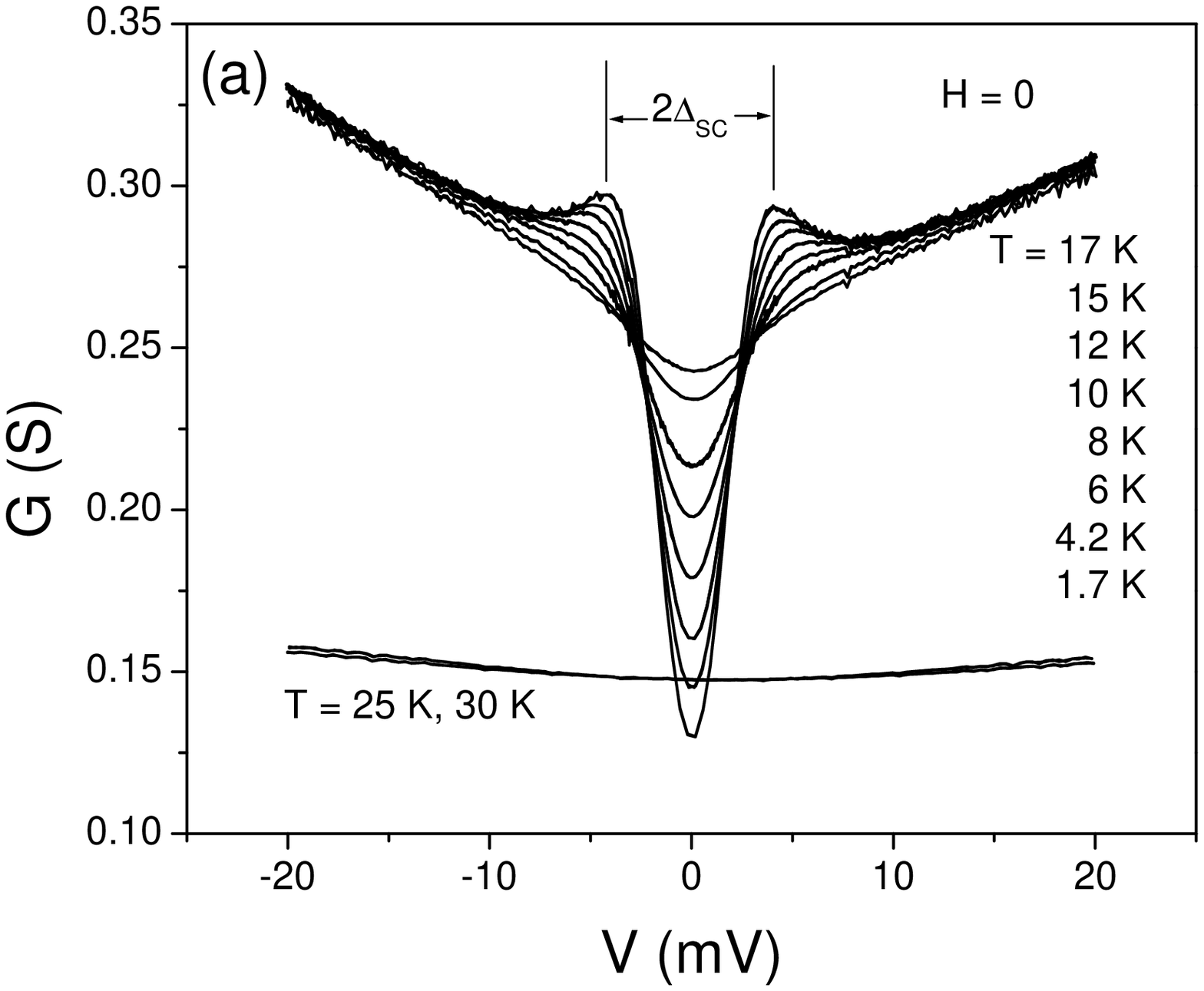,width=8.0cm,height=6.0cm,clip=}
}
\centerline{
\psfig{figure=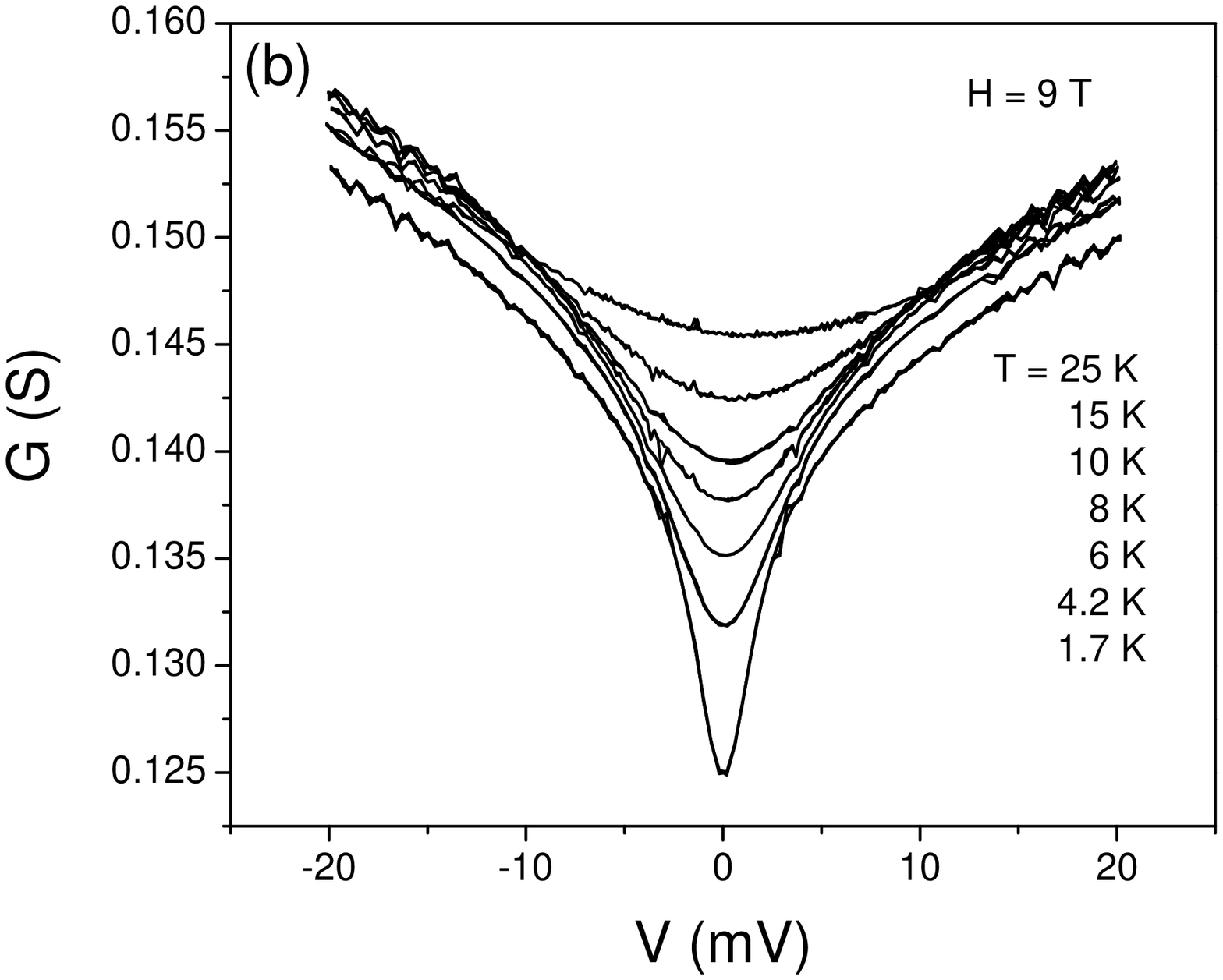,width=8.0cm,height=6.0cm,clip=}
}
\caption{(a) The $G-V$ curves for the break junction
between silver
and PCCO ($x=0.15$). The superconducting gap is marked. The curves above
$T_c$ are shifted due to the resistance of the film in series with the
junction. (b) $G-V$ curves for the same junction in (a) in a magnetic field
of 9 tesla applied parallel to the $c$-axis. A prominent gap in the normal
state is seen at low temperatures.}
\end{figure}
the data using the break junction method.
The superconducting gap is clearly visible and results in a bunching
of states near $\pm\Delta_{SC}$ (coherence peaks). 
The value of $\Delta_{SC}$ is about 5 meV which is similar to the
values reported in earlier tunneling studies ~\cite{zasadzinski,alff2}.
The conductance
($G=dI/dV$)
within the gap voltage does not go to zero because of two 
possible reasons:
1) the junctions have more than one conduction
channel i.e. are not tunnel junctions but ballistic point contact
junctions and 2) the likely $d$- wave symmetry of the SC gap. The effect 
of the first point has been discussed in detail in the classic paper
by Blonder, Tinkham and Klapwijk (BTK) ~\cite{btk} for the case of
isotropic $s$-wave superconductors. In the
BTK analysis the junction was parametrized
by a quantity $Z$ which depends on the scattering at the interface
between the metal and the superconductor. High transparency junctions have
$Z \sim 0$ and tunnel junctions have $Z \gg 1$. For superconductors
with an anisotropic gap a modified BTK model has been described in
ref. ~\cite{kashiwaya1}. A detailed fitting of our data to such
models is in progress. 
\begin{figure}
\centerline{
\psfig{figure=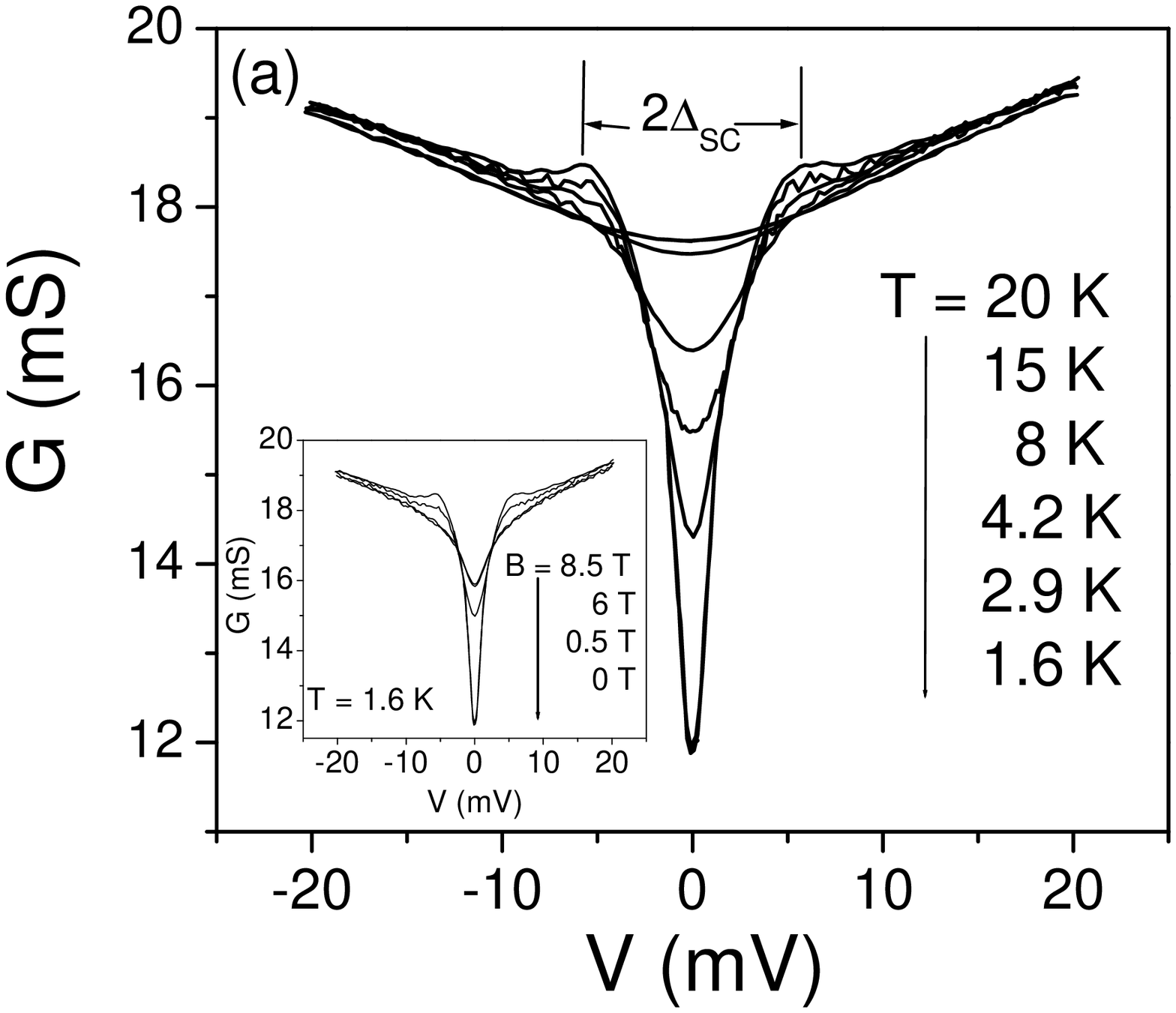,width=8.0cm,height=6.0cm,clip=}
}
\centerline{
\psfig{figure=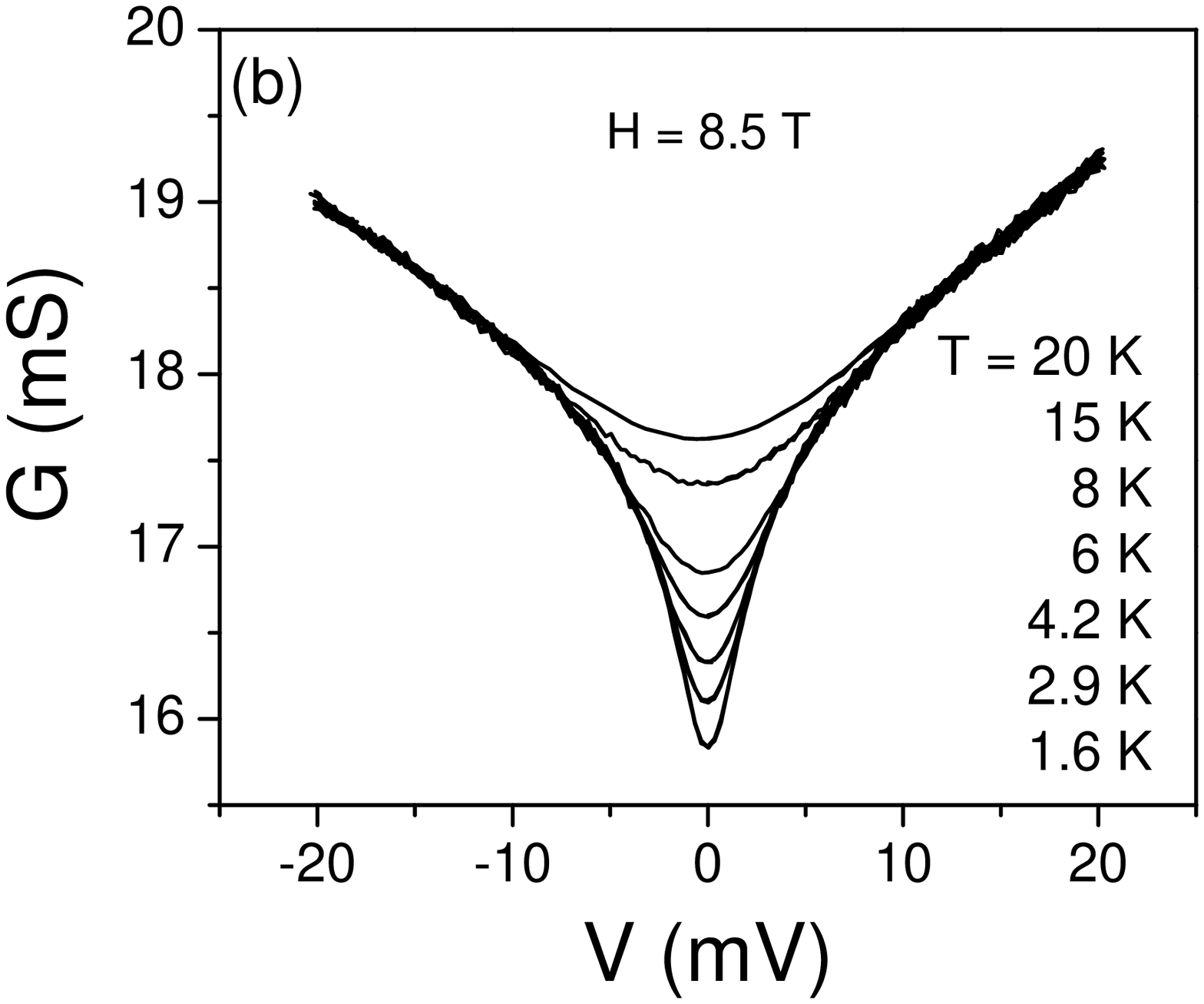,width=8.0cm,height=6.0cm,clip=}
}
\caption{(a) The $G-V$ curves for the pressed indium junction with PCCO
($x=0.15$). The superconducting gap is marked. The inset shows the
effect of the indium gap on the $G-V$ curves. (b) The $G-V$ curves for
the same junction in (a) in a magnetic field of 8.5 tesla applied parallel
to the $c$-axis. The normal state gap is clearly visible.}
\end{figure}
However in the following sections we will use these
ideas qualitatively. Fig. 1a also shows the variation of the
tunneling spectra with temperature. This 
variation is due to a combined effect
of the thermal smearing of the spectra and the variation in the 
density of states (DOS)
of the SC with temperature. The $\Delta_{SC}$ does not change noticeably
with temperature. The gap structure disappears above $T_c$ and the
background conductance of the junction is observed for $T$=25 K.

When a magnetic field $H > H_{c2}$
is applied to this tunnel junction at $T =$ 1.7 K ($T \ll T_c$), 
the SC gap is no longer observed as expected (i.e. the coherence peaks
disappear), {\em but surprisingly} the gap at the $E_F$ persists. However
the shape of this gap as a function of energy is quite different (fig. 1b)
from the SC gap. 
We call this unexpected gap the normal state gap (NSG), since resistivity
measurements at this magnetic field show that the PCCO film is in the 
normal state. 
Before we further analyze this NSG
we present data which show that the
observation of this NSG does not depend on the type of 
\begin{figure}
\centerline{
\psfig{figure=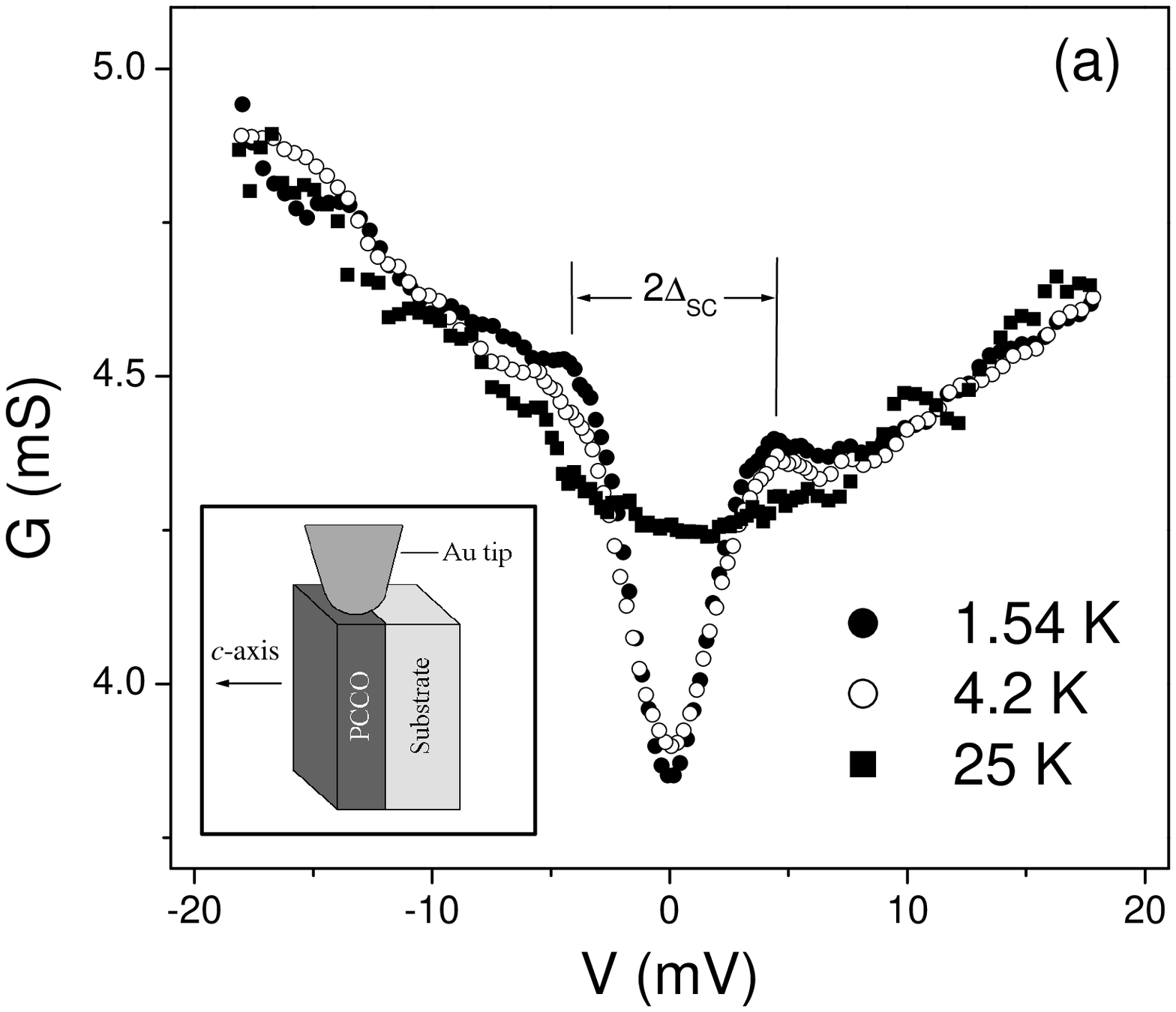,width=8.0cm,height=6.0cm,clip=}
}
\centerline{
\psfig{figure=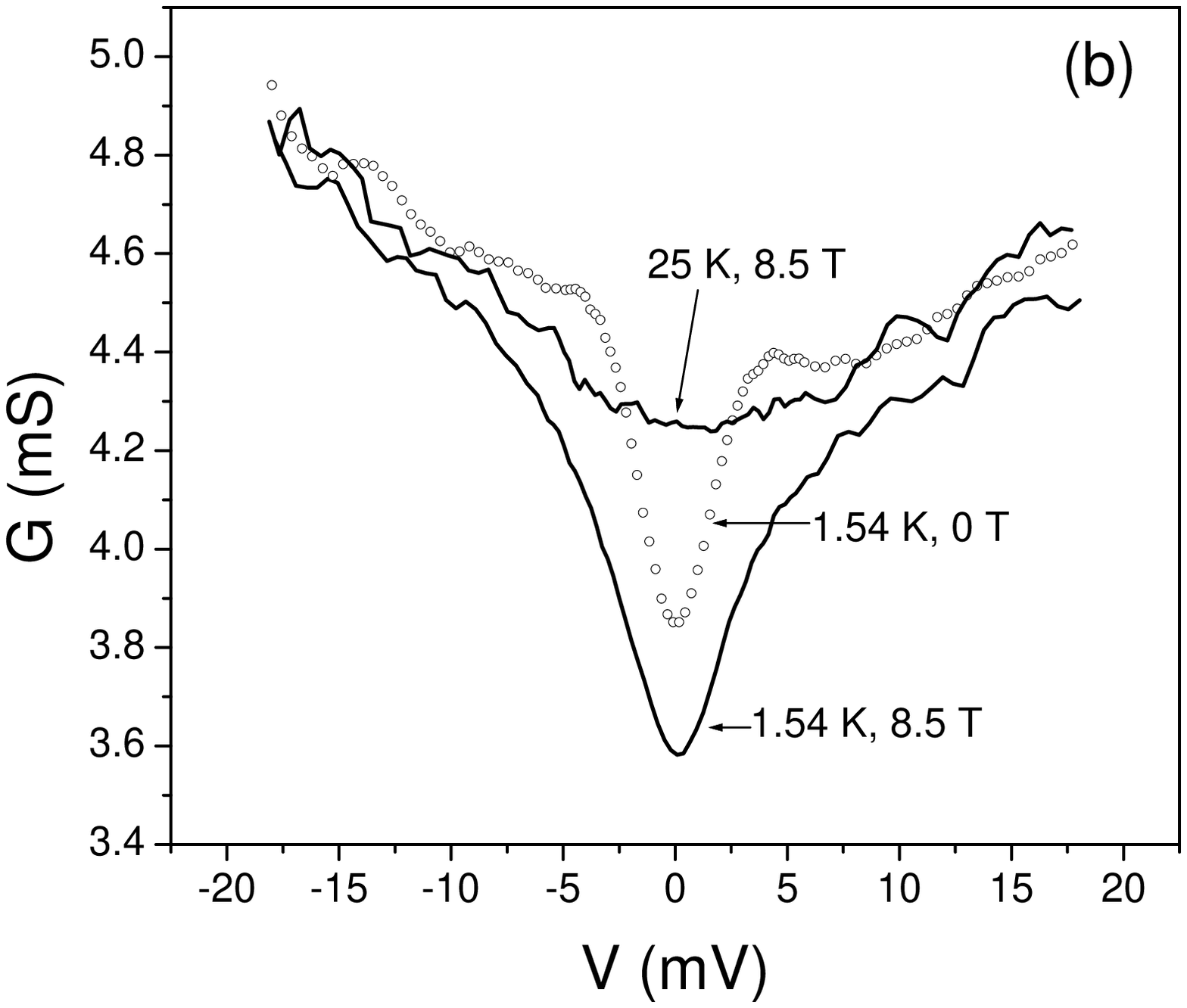,width=8.0cm,height=6.0cm,clip=}
}
\caption{(a) The $G-V$ curves for the point contact junction between PCCO
($x=0.15$) and gold. The superconducting gap is marked for the curve
taken at 1.54 K.
The inset shows the setup for the formation of the point contact
junction between PCCO and gold for tunneling into the $a-b$ plane.
(b) The normal state gap is seen when a field of 8.5 T is applied
parallel to the $c$-axis.}
\end{figure}
tunnel junction.
Fig 2a shows data for a pressed junction using an indium counterelectrode.
The indium was used to check the quality of the junction. The 
contribution of the indium density of states is visible in the
zero field data, since when a field of about 5000 G was 
applied to make the
indium electrode normal there is a large increase in the
zero bias conductance which indicates that the indium has become a 
normal metal with no superconducting gap (inset fig. 2a).
In the superconducting state of indium the $G$ at zero bias is not
zero which shows that this junction is not in the tunneling limit.
We find
$Z \approx 0.7$ according to a rough estimate made from the ratio of
the conductance at 20 mV ($G_{20}$) and at 0 V ($G_0$) ~\cite{btk}.
For fields above 5000 G 
the contributions to the tunneling
spectra will be dominated by the DOS features of PCCO. For an applied
field of 8.5 T the NSG is clearly seen for this junction too (fig. 2b). 
In fig. 3a we show the tunneling data obtained from point contact tunnel 
junctions formed between a gold tip and a PCCO thin film.
The zero field data clearly shows the
\begin{figure}
\centerline{
\psfig{figure=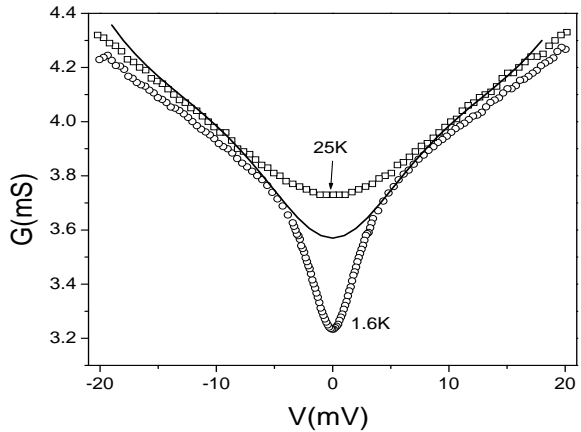,width=8.0cm,height=6.0cm,clip=}
}
\caption{The effect of thermal smearing on the normal state
gap in the tunneling spectra. The
normal state gap at 1.6 K (open circles)
evolves to the solid line at 25 K
due to thermal
smearing (calculated using eqn. 1).
This calculated curve at 25 K does not match
the experimental curve at 25 K (open squares). Data shown is for
$x=0.15$ composition using the break junction method, in a field of 8.5 T.}
\end{figure}
gap feature for
superconducting PCCO. The temperature dependence of the tunneling spectra 
is similar to that observed for the other two types of junctions.
On the application of a magnetic field of 8.5 T
we observe the formation of the NSG again (fig. 3b). 
From these data we conclude
that this anomalous NSG is present irrespective of the kind of
tunnel junction. 

We now discuss the nature of this NSG in more detail.
First we look at the behavior of this gap as the temperature is increased
upto and beyond $T_c$. The $T_c$ for our optimally doped PCCO films is 
about 21 K. We see that the NSG becomes indistinguishable from the
background for $T > T_c$ (figs. 1b, 2b and 3b). This could be interpreted
as evidence for the
NSG being due to some residual superconductivity
(even for $H > H_{c2}$) which goes away for $T > T_c$. 
However we have to consider that the width of the NSG is $\sim$ 7 meV and
a temperature of 25 K can smear out
features of about 5 meV in the tunneling spectra. To check if the 
NSG closes around $T = T_c$ just due to thermal smearing we have 
made a simple estimate of
the role of thermal smearing in the temperature evolution of the
$G vs. V$ curves using the following equation for the tunneling
current ~\cite{wolf}:
\begin{eqnarray}
I(V,T) & = & c\int^{\infty}_{-\infty}n\left(E+\frac{eV}{2}\right) \nonumber\\
& &\times\left[f\left(E+\frac{eV}{2}, T\right)
-f\left(E-\frac{eV}{2}, T\right) \right] dE
\end{eqnarray}
where $n(E)$ is the DOS of the PCCO sample at zero temperature,
$f(E, T)$ is the Fermi function at a temperature $T$, and V is the
junction bias. For $n(E)$ we have taken the $G(V) vs. V$ curve at
1.6 K and smeared it using eqn. 1 for a temperature $T$. We have
assumed that the density of states of the counterelectrode
gold is a constant 
for the relevant energy range
and so is the tunneling barrier for this range
of energies both of which are included in the constant $c$. The results
of these calculations are shown in fig. 4. It  shows 
that thermal smearing 
alone cannot explain the disappearance of the NSG for $T \sim$ 20 K.
{\em This means that the NSG has a significant 
thermal dependence of its own.}
With this analysis we have shown that 
although the thermal smearing of the tunneling
spectra has a significant influence on their evolution with temperature
the NSG does indeed
become smaller with increasing temperature. This thermal
dependence of the NSG could be due to residual SC, or it could be that
there is no residual SC and the NSG has its own thermal dependence. 

To study the possible contribution to the NSG of residual SC at the
interface and/or the tunnel barrier,
we formed a high transparency point contact junction between a 
gold tip and an $x = 0.13$ thin film of PCCO.
Fig. 5a shows the $G-V$ data at various applied fields
for such a junction, at 1.6 K. The peak formed at
zero bias (for $H < H_{c2}$) 
is due to Andreev reflection which is observed in high
transparency junctions. At $H=0$ 
the width of the feature is about $2\Delta_{SC}$ ($\sim$ 10 meV) and
$G(0)/G_N$ is 1.62 (fig. 5a inset), 
where $G_N$ is the value of $G(V)$ for $eV > \Delta$.
These are clear signatures of Andreev reflection (AR) occurring at the
interface between gold and PCCO. This also shows that for this junction
$Z \sim 0$. We chose the $x = 0.13$ composition
because it has a low $T_c$ of 12 K and therefore we get a good temperature
range to operate in before thermal smearing washes out the features in the
$G(V) vs. V$ curves. When a magnetic field is applied on this junction,
the AR feature reduces in height and also shrinks in width and when the 
superconductivity is destroyed, disappears completely. This
is a clear indication that there is no residual SC at the
interface in our $G-V$ curves. However, the NSG is clearly
visible! In fact for intermediate values of $H$ the
normal state density of states reveals itself at higher $V$ while at
the same time there is peak at zero bias due to Andreev reflection,
as shown in fig 5b.
Since this is a high transparency junction, 
the junction barrier has negligible effect on the shape of the curve.
This is clear from fig. 5a since for $H = 0$ 
the $G-V$ curve is almost
flat (independent of $V$) for $eV > \Delta$. 

The other test to
check if the NSG is truly a feature of the normal state is to
increase the temperature beyond $T_c$ and observe the evolution of the 
$G(V) vs. V$ curves. Fig. 5c shows that removing SC by increasing 
the temperature has
an evolution similar to that of increasing $H$ beyond $H_{c2}$
(for $T < T_c$). 
However, the presence of the NSG is less
clear for $T > T_c$ due to thermal smearing.
This strongly suggests that
the NSG is indeed a normal state feature. We should add
here that the high transparency junctions are not tunnel junctions
but point contact junctions with more than one channel for conduction.
However we are still in the ballistic region i.e. the junction size,
given by a radius $a$,
is smaller than the elastic mean free path of the electron, $l_e$
(the Sharvin 
\begin{figure}
\centerline{
\psfig{figure=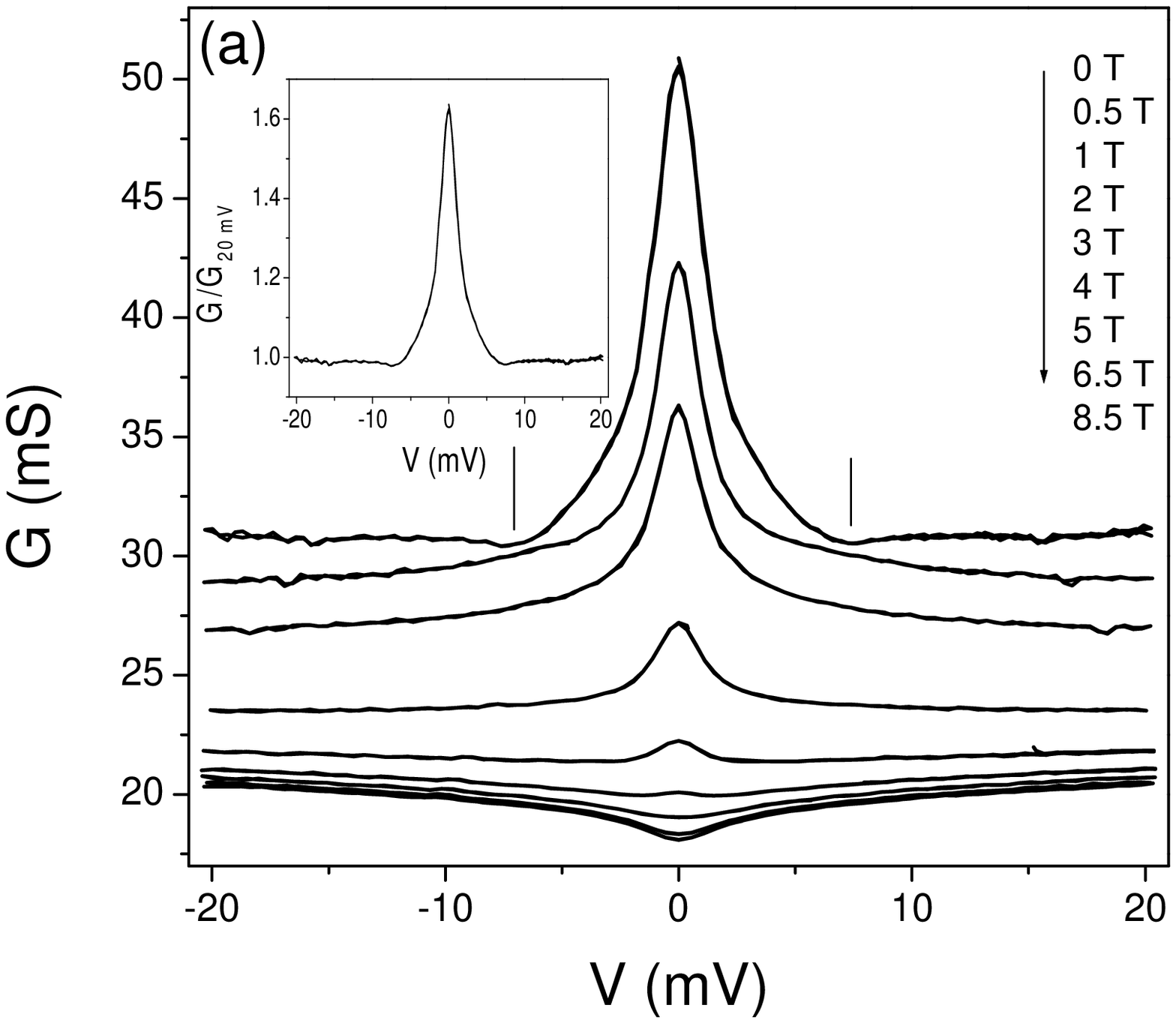,width=8.0cm,height=6.0cm,clip=}
}
\centerline{
\psfig{figure=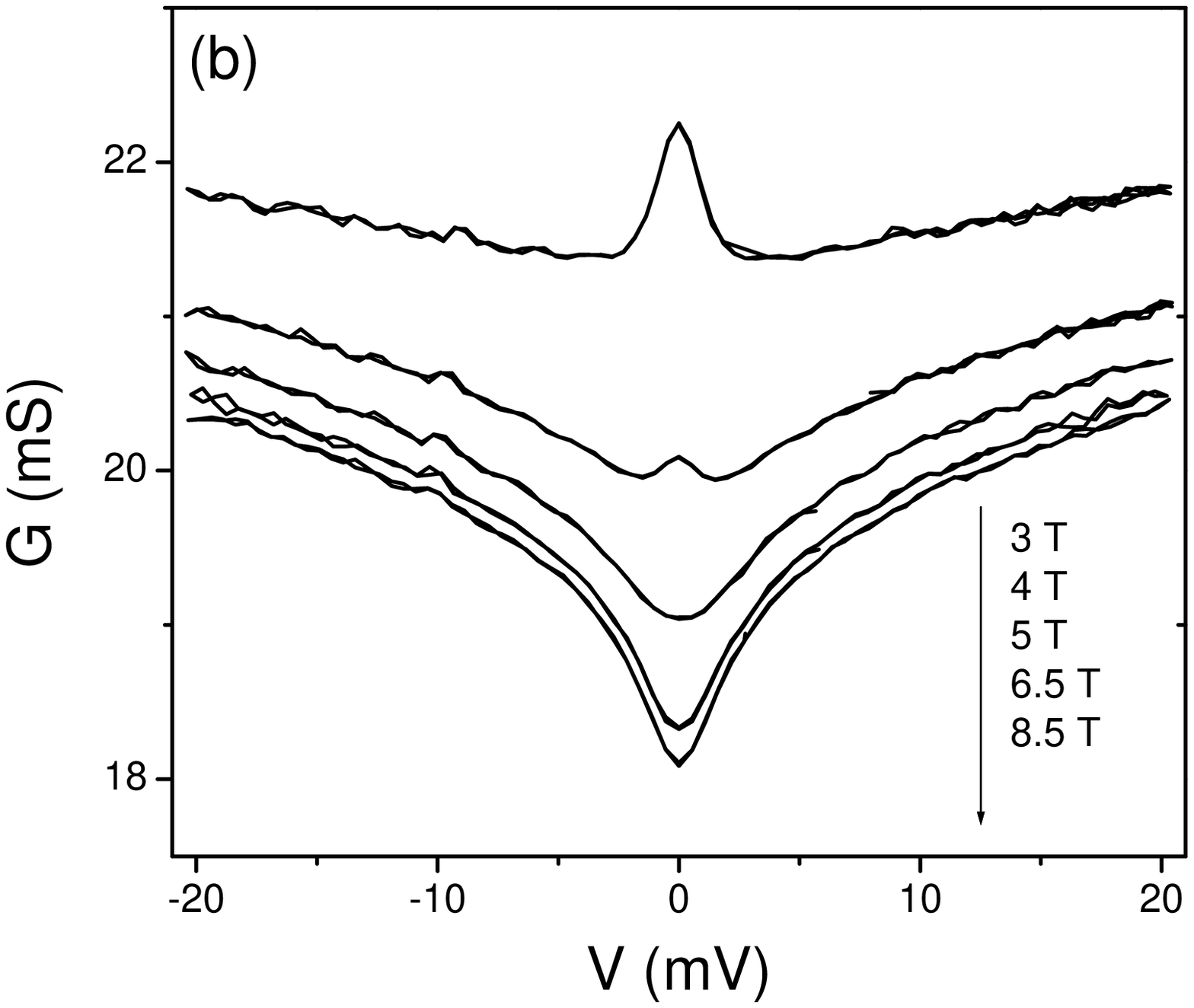,width=8.0cm,height=6.0cm,clip=}
}
\centerline{
\psfig{figure=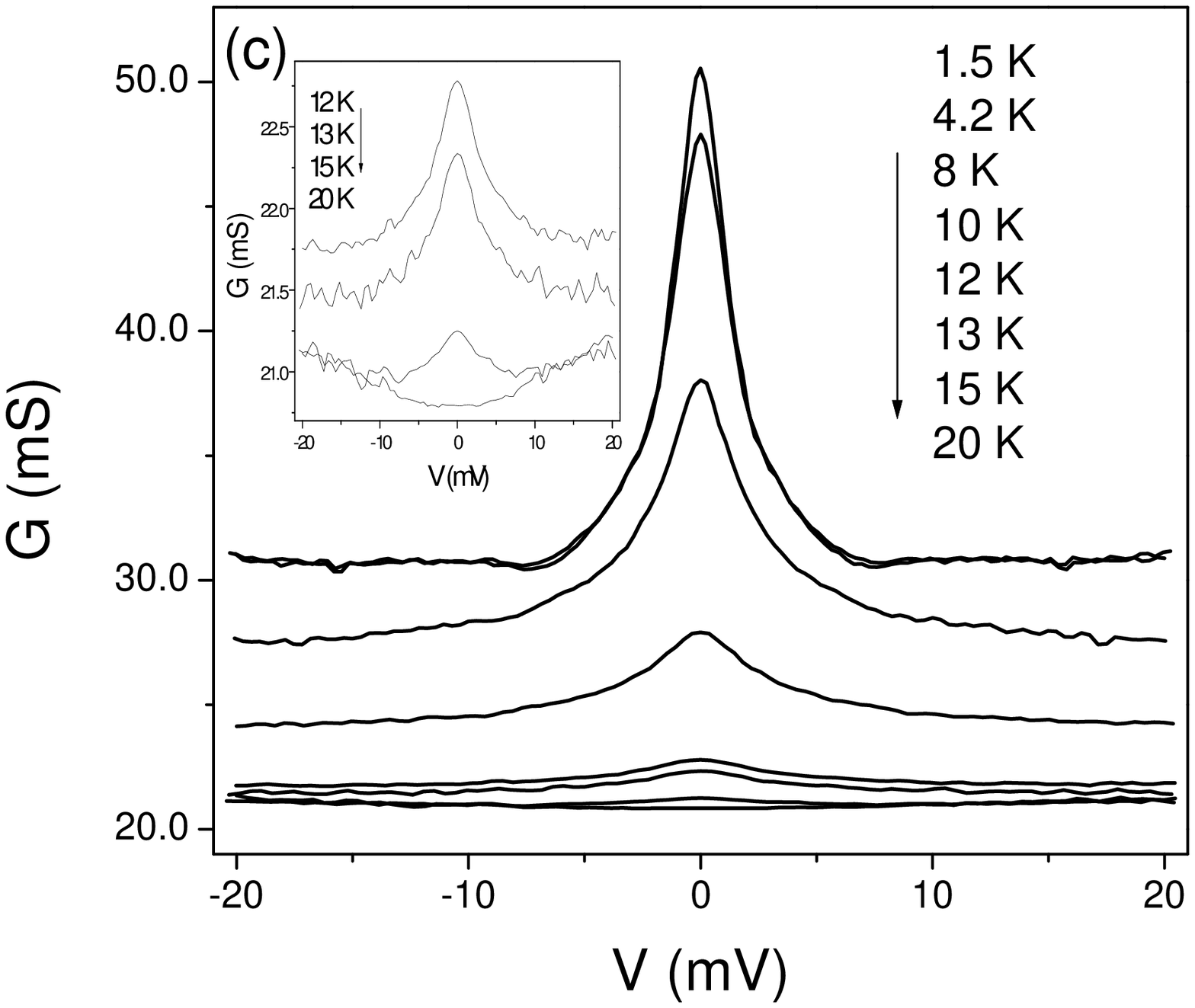,width=8.0cm,height=6.0cm,clip=}
}
\caption{(a)
The $G-V$ curves for a high transparency ($Z \sim 0$) point contact
junction between PCCO ($x=0.13$) and gold. A peak due to Andreev reflection
(AR)is seen near zero bias.  The inset shows the increase in $G$ by a factor
of  1.62 at zero bias compared to the $G$ at 20 mV. When a field is applied,
the AR peak shrinks both in height and width and disappears for fields
$\ge$ 5 T. (b) An enlarged view of the $G-V$ curves at high magnetic fields.
As the AR peak shrinks and disappears the normal state DOS is seen, even
when there is a peak at zero bias, for $\mid V\mid > \Delta_{SC}$. The
normal state gap is again seen for $H \ge$ 5 T. (c) The variation of the
$G-V$ curves with temperature. The normal state $G-V$ curve is seen at 20 K.
The inset shows an enlarged view of the $G-V$ curves at high temperatures.}
\end{figure}
\begin{figure}
\centerline{
\psfig{figure=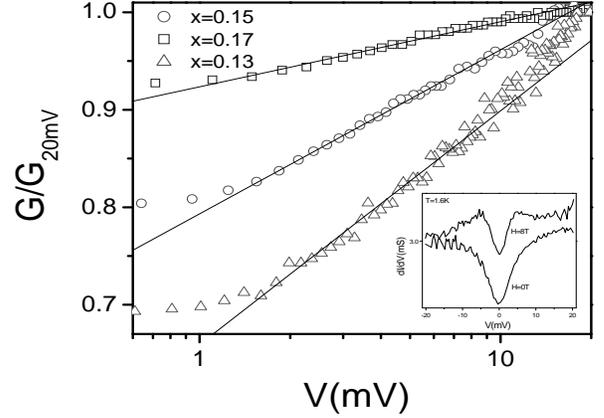,width=8.0cm,height=6.0cm,clip=}
}
\caption{A plot of $G~ vs. logV$ in the normal state of underdoped ($x=0.13$,
$T=1.61$ K),
overdoped($x=0.17$, $T=1.6$ K)
and optimally doped ($x=0.15$, $T=1.54$ K) compositions of PCCO. Data
for positive bias is shown. The solid lines are linear fits to the data for
a certain voltage range. The inset shows tunneling spectra for PCCO ($x=0.17$)
at 1.6 K,
in the superconducting ($H=0$) and normal ($H=8.5$ T) states.}
\end{figure}
limit). We know this must be the case because at 
$T \sim$ 20 K,
$G(V)$ has a minimum at $V = 0$ ~\cite{srikanth1}.  
If the junction were in the diffusive limit ($a > l_e$)
then the curvature of the
tunneling spectra in the normal state would change and there 
would be a maximum in the $G-V$
curve at $V = 0$. Hence the features
observed in the $G-V$ curves in the normal state do reflect the
DOS features of the normal state.
From this experiment we thus draw two
important conclusions: 1) the NSG is 
definitely a normal state feature and
2) it can be observed for $H > H_{c2}$ as well as for $T > T_c$.

What is the origin of this NSG? 
Such features in the density of states have been observed
before in various systems like disordered metallic films ~\cite{imry} and
perovskite oxides like LaNiO$_{3-\delta}$ ~\cite{srikanth2}
using tunneling spectroscopy. 
The appearance of these features in the density of states of these materials
is due to
electron-electron Coulomb interactions which are enhanced due to
disorder in the materials. Such effects lead to 
anomalies in the DOS, $n(E)$,
near $E_F$. For three-dimensional systems this anomaly is of the form
$n(E)=n(0)[1+\sqrt{E/\Delta}]$
where $\Delta$ is called the correlation gap.
For two-dimensional systems the electron-electron interaction leads
to decrease in the DOS near $E_F$ given by $\delta n/n \sim -ln(E/E_0)$
~\cite{altschuler}.
These anomalies are the precursors of the insulating phase
``Coulomb gap'' which is formed at $E_F$ when the disorder is increased
and the states near $E_F$ are localized. 

To check if such effects are
the reason for the formation of the NSG we look at the
properties of the NSG is more detail.
Fig 6 shows a plot of the $G(V)$ as a function of $log V$ ($V > 0$) taken at
$T = 1.6$ K using a gold and PCCO point contact junction.
$G(V)$ is a linear function of $log V$ for a large range of V (for $V
< 1$ mV, $G(V)$ deviates from the straight line due to thermal smearing).
This behavior follows closely the abovementioned
signature of a 2D correlation gap ~\cite{imry,butko}. 
In ref. ~\cite{imry} it is shown that as the sheet resistance of the
film is increased, the logarithmic anomaly at zero bias becomes deeper.
We want to verify if a similar trend is observed in the normal state
of PCCO as in the case of disordered metal films.
The sheet resistance in the normal state of PCCO increases as $x$ is 
lowered ~\cite{pat1}.
Moreover, the field induced normal state in PCCO
has a metallic behavior for overdoped samples ($x > 0.15$)
and the resistivity shows an upturn at low temperatures for optimally
doped and underdoped samples ($x \le 0.15$) ~\cite{pat1}. A 
suggested mechanism for this
transition was the onset of weak localization below a certain doping.
This transition suggests an increase in the electron-electron
interactions as $x$ is decreased. If our hypothesis about effect of
interactions on the DOS is correct then the tunneling data should
reflect this transition. The inset of 
fig. 6 shows the point contact tunneling data for an $x=0.17$ sample of
PCCO ($T=1.6$ K). 
At zero field the superconducting gap is clearly seen. At 8.5 T
the SC gap gives way to the NSG. The NSG in this case is smaller than
the one found for $x=0.15$. Fig. 6 also shows the 
$G_{norm}-log V$ plots
for the $x= 0.17$ (at $T=1.6$ K) 
and $x = 0.13$ (at $T=1.61$ K) samples. The deepening of the NSG
with lowering of $x$ is clearly indicated by the increase of the
slope of the $G_{norm}~vs. ~logV$ curves and
is similar to the trend observed in metallic films with 
different amounts of disorder ~\cite{imry,butko}. Therefore our tunneling
data shows that in the underdoped region there are strong
electron-electron interactions which 
results in a larger NSG. As the doping
is increased the effect of the interactions is lowered and a 
trend towards a more
conventional Fermi liquid like normal state is obtained as seen from a 
much smaller NSG. It has been conjectured that
for some value of $x$ there is quantum critical transition in the
normal state of cuprate superconductors
~\cite{sachdev}. 
The gradual disappearance of the NSG is an indication that the physical
properties of the cuprates are going through a gradual crossover with
increasing $x$, probably due to their proximity to this quantum critical
point.

The data discussed above show that there is a NSG in the normal 
state of
PCCO. This gap shows the properties of a 2D correlation gap and
is related to the position of the material in the PCCO phase
diagram i.e. it depends on the doping. The pseudogap (PG)
is a gap in 
the normal state of cuprate superconductors and it also depends
on the doping ~\cite{timusk}. In fact the other similarity between
the NSG observed here and the PG is that the gap becomes smaller for
the overdoped compounds.  
But is our observed NSG the same as the pseudogap (PG)?
In ref. ~\cite{alff1} the authors claim that the NSG they
observe in PCCO and NCCO is the pseudogap and they also 
estimate field values at which this NSG is suppressed.
However there is serious discrepancy in the energy scale of the
pseudogap observed in ref. ~\cite{alff1} and in our
experiments with that observed in 
other experiments.
Optical conductivity data on LSCO
(a hole-doped cuprate superconductor similar to PCCO)
show a PG of about 500 cm$^{-1}$ i.e. about 62 meV
~\cite{startseva}. New optical conductivity data on e-doped
NCCO gives a large pseudogap width of about 600 cm$^{-1}$ which
survives even above room temperature ~\cite{basov2}. 
Recent ARPES data
also shows evidence of a pseudogap larger than 50 meV
~\cite{armitage}. In
non-superconducting NCCO pseudogaps of about 200 meV have been
observed from optical conductivity
measurements ~\cite{tokura}. These gaps are at least
an order of magnitude larger than the NSG we observe. For high $T_c$
compounds like BSCCO the PG observed from the tunneling spectra
agree quite well with other experiments like ARPES and optical
conductivity ~\cite{timusk}. Hence, it is unlikely
that the NSG we found in our tunneling experiments in
PCCO and the pseudogap observed in other 
experiments on PCCO and NCCO are the same. 

There is only a small variation of the NSG with 
magnetic field in our experiment and for the maximum field values
that we applied ($\sim$ 9 T) the NSG was still prominent. The 
small variation that we see with the magnetic field is most likely
due to the change in $e-e$ correlations on the application of a 
magnetic field ~\cite{patfilm3}.
However, further experiments are necessary to determine
accurately the variation of the NSG with $T$ and $H$.

\section{Summary and Conclusions}

In this paper we have shown that the density of states of the normal
state of PCCO shows a depletion of states near the Fermi level $E_F$.
This normal state gap can be treated as a correlation gap which is
formed due to electron correlations. Such gaps have been seen in
disordered metals using tunneling spectroscopy. We have argued that
this gap is probably not the pseudogap mainly due to a difference
of an order of magnitude in their values. We have shown that the 
nature of the NSG depends on the doping i.e. it depends on the
position in the phase diagram. We conjecture that this NSG is
linked with a quantum phase transition which most probably
occurs in the normal 
state of the cuprates for a certain value of hole or electron doping.

\acknowledgements
P. F. acknowledges the support of the Canadian Institute of Advanced
Research and the Foundation Force of the Universite de Sherbrooke.
We thank Hamza Balci and Z. Y. Li for some of the thin film samples.
This work was supported by NSF DMR 97-32736


\begin{references}
\bibitem{suzuki} M. Suzuki and T. Watanabe, Phys. Rev. Lett. {\bf 85},
4787 (2000)
\bibitem{timusk} T. Timusk and B. Statt, Rep. Prog. Phys. {\bf 62}, 61
(1999)
\bibitem{sato} T. Sato {\em et al.}, Phys. Rev. Lett. {\bf 83}, 2254 
(1999)
\bibitem{startseva} T. Startseva {\em et al.}, Phys. Rev. B {\bf 59},
7184 (1999)
\bibitem{alff2} L. Alff {\em et al.}, Phys. Rev. B {\bf 58}, 11197 (1998)
\bibitem{patgap} J. David Kokales, Patrick Fournier, Lucia V. Mercaldo,
Vladimir V. Talanov, Richard L. Greene, and
Steven M. Anlage, Phys. Rev. Lett. {\bf 85} 3696 (2000)
\bibitem{prozorov} R. Prozorov, R. W. Giannetta, P. Fournier and
R. L. Greene, Phys. Rev. Lett. {\bf 85}, 3700 (2000)
\bibitem{kirtley} C. C. Tsuei and J. R. Kirtley, Phys. Rev. Lett.
{\bf 85}, 182 (2000)
\bibitem{pat1} P. Fournier, P. Mohanty, E. Maiser, S. Darzens, T. Venkatesan,
C. J. Lobb, G. Czjzek, R. A. Webb, and R. L. Greene, Phys. Rev. Lett.
{\bf 81}, 4720 (1998)
\bibitem{ourpap} Amlan Biswas, P. Fournier, V.N. Smolyaninova, 
J. S. Higgins, H. Balci, R.C. Budhani, and R.L. Greene,  APS March
meeting proceedings, abstract no. A27.006 (2001)
\bibitem{alff1} S. Kleefisch, B. Welter, A. Marx, L. Alff, R. Gross,
and M. Naito, Phys. Rev. B {\bf 63}, 100507(R) (2001)
\bibitem{armitage} N. P. Armitage {\em et al.}, Private Communication
\bibitem{basov2} E. J. Singley, D. N. Basov, K. Kurahashi, T. Uefuji,
and K. Yamada, cond-mat/0103480 
\bibitem{huang} Q. Huang {\em et al.}, Nature {\bf 347}, 369 (1990)
\bibitem{yamamoto} H. Yamamoto, M. Naito and H. Sato, Phys. Rev. B
{\bf 56}, 2852 (1997)
\bibitem{ekino} T. Ekino, T. Doukan and H. Fujii, J. Low. Temp. Phys.
{\bf 105}, 563 (1996)
\bibitem{patfilm1} E. Maiser {\em et al.} Physica (Amsterdam) {\bf
297C} 15 (1998)
\bibitem{patfilm2} J. -L. Peng {\em et al.} Phys. Rev. B {\bf 55}, R6145
(1997) 
\bibitem{zasadzinski} Q. Huang, J. F. Zasadzinski, N. Tralshawala,
K. E. Gray, D. G. Hinks, J. L. Peng, and R. L. Greene, Nature {\bf
347}, 369 (1990)
\bibitem{btk} G. E. Blonder, M. Tinkham, and T. M. Klapwijk, Phys.
Rev. B {\bf 25}, 4515 (1982)
\bibitem{kashiwaya1} S. Kashiwaya {\em et al.}, Phys. Rev. B {\bf 51},
1350 (1995)
\bibitem{wolf} E. L. Wolf, {\em Principles of Electron Tunneling
Spectroscopy} (Oxford University Press, New York, 1985)
\bibitem{srikanth1} H. Srikanth and A. K. Raychaudhuri, Phys. Rev. B
{\bf 46}, 14713 (1992)
\bibitem{imry} Y. Imry and Z. Ovadyahu, Phys. Rev. Lett. {\bf 49}, 841
(1982)
\bibitem{srikanth2} A. K. Raychaudhuri, K. P. Rajeev, H. Srikanth, 
and N. Gayathri,
Phys. Rev. B {\bf 51}, 7421 (1995)
\bibitem{altschuler} B. L. Altshuler, A. G. Aronov, and P. A. Lee,
Phys. Rev. Lett. {\bf 44}, 1288 (1980)
\bibitem{butko} V. Yu. Butko, J. F. DiTusa, and P. W. Adams, Phys.
Rev. Lett. {\bf 84}, 1543 (2000)
\bibitem{sachdev} S. Sachdev, Science {\bf 288}, 475 (2000)
\bibitem{tokura} Y. Tokura, International
Workshop on Novel Quantum Phenomena in Transition Metal
Oxides, Conference Proceedings, pp. 61 (2000)
\bibitem{patfilm3} P. Fournier, J. Higgins, H. Balci, E. Maiser, 
C. J. Lobb, and R. L. Greene, Phys. Rev. B {\bf 62} R11993 (2000) 
\end{references}
\end{document}